\documentclass[10pt,letterpaper]{article}
\usepackage{opex3}
\usepackage{epsfig,graphics}
\usepackage{graphicx}
\usepackage{amsmath}
\usepackage{amsfonts}
\begin{document}

%%%%%%%%%%%%%%%%%% title page information %%%%%%%%%%%%%%%%%%
\title{Controlled Manipulation of Mode Splitting in an Optical Microcavity by Two Rayleigh Scatterers}

\author{Jiangang Zhu, \c{S}ahin Kaya \"{O}zdemir, Lina He and Lan Yang$^{*}$}

\address{Department of Electrical and Systems Engineering, Washington University, St. Louis, Missouri 63130, USA}

\email{yang@seas.wustl.edu} 
\begin{abstract}
We report controlled manipulation of mode splitting in an optical microresonator coupled to two nanoprobes. It is demonstrated that, by controlling the positions of the nanoprobes, the split modes can be tuned simultaneously or individually and experience crossing or anti-crossing in frequency and linewidth. A tunable transition between standing wave mode and travelling wave mode is also observed. Underlying physics is discussed by developing a two-scatterer model which can be extended to multiple scatterers. Observed rich dynamics and tunability of split modes in a single microresonator will find immediate applications in optical sensing, opto-mechanics, filters and will provide a platform to study strong light-matter interactions in two-mode cavities.
\end{abstract}

 % REPLACE WITH CORRECT OCIS CODES FOR YOUR ARTICLE

%%%%%%%%%%%%%%%%%%%%%%% References %%%%%%%%%%%%%%%%%%%%%%%%%

%%%%%%%%%%%%%%%%%%%%%%%%%%  body  %%%%%%%%%%%%%%%%%%%%%%%%%%

\section{Introduction}
Whispering-gallery-mode (WGM) optical microresonators with ultra-high quality factors and microscale mode volumes are of interest for a variety of scientific disciplines ranging from fundamental science to engineering physics. Significantly enhanced light-matter interactions \cite{Microcavity} make WGM resonators remarkably sensitive transducers for detecting perturbations in and around the resonator, e.g., virus/nanoparticle detection at single particle resolution \cite{label-free,virus,Zhu} and ultrasensitive detection of micromechnical displacement \cite{radiationpressure}. Moreover, the coupling of optical and mechanical modes mediated by enhanced radiation pressure within the microresonator provides a superb platform to study parametric oscillation instabilities and radiation pressure induced cooling of mechanical modes \cite{optomechanics}. In addition, WGM microresonators with asymmetry (e.g., induced by structural deformations etc.) have been useful for investigating the correspondence between quasieigenstates and associated classical dynamics in mesoscopic systems \cite{ExceptionalPoint}. Level crossing have been demonstrated in microtoroids by tuning the microtoroid aspect ratio \cite{levelcrossing}.

One interesting phenomenon associated with enhanced light-matter interactions is the splitting of the initially degenerate cavity modes in the strong coupling regime \cite{Weiss,KippenModalCoupling,Mazzei,Deych}. Mode splitting manifests itself as a doublet (two resonances) in the transmission spectrum of the resonator. Here we show, for the first time, that by tuning the coupling strength using nanoprobes or subwavelength scatterers, the split modes can be manipulated individually or together forming a tunable two-mode microcavity. We  demonstrate that the modes can be tuned to cross or anti-cross in frequency and linewidth. Particularly interesting experimental observation is the tunable transition between standing wave mode (SWM) to travelling wave mode (TWM) and vice versa. In order to explain the observed rich dynamics, we develop a model which takes into account the interactions between WGMs and finite number of sub-wavelength scatterers. Results of the numerical simulations based on this model explains and confirms the experimental observations as well as suggests a rather surprising dynamics, i.e., the two split modes can be made to cross each other twice in linewidth with frequencies undergoing anticrossing in one of the linewidth crossing region. All the observed phenomena imply there are exceptional points in this cavity-scatterers system.

\section{Experiments, Theoretical Model and Discussions}
\begin{figure}[htbp]
\centering\includegraphics[width=10cm]{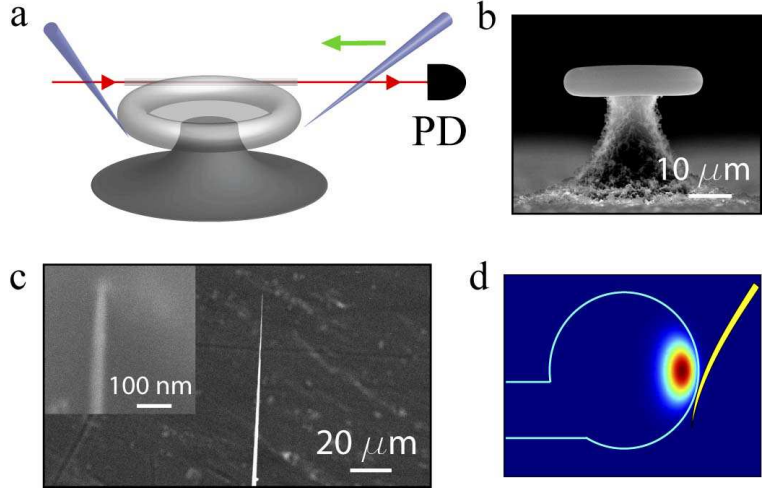}
\caption{(a) Schematics of the experiment showing a fiber taper coupled to a microtoroid, and two nano-fiber tips introduced into the mode volume. Transmission spectra are captured by a photo detector (PD). (b) SEM image of a microtoroid. (c) SEM image of a fiber tip. Inset shows the enlarged image of the tip. (d) Cross-section of a microtoroid ring showing the position of a fiber tip in the field of a WGM.}\label{fig1}
\end{figure}

Figure \ref{fig1} depicts the schematics of our experimental scheme which is composed of a WGM silica microtoroid  resonator coupled to two nanoprobes prepared by heat-and-pull of optical fibers on a hydrogen flame followed by buffered HF etching. To couple light in and out of the resonator, a fiber taper is used. Positions of the nanoprobes and the fiber taper are finely controlled by 3D translational stages. We place the first nanoprobe in the resonator mode volume and fix its position when mode splitting is observed. We denote individual resonances of the doublet as $\omega^{-}$(lower freuqency) and $\omega^{+}$(higher frequency) modes with corresponding linewidths $\gamma^{-}$ and $\gamma^{+}$ ($\gamma^{-}>\gamma^{+}$), respectively. Then the second probe is introduced. This probe bends down and slides along the surface vertically as it contacts the rim of the microtoroid. Due to cone-like shape of the tip, vertical movement gradually increases its diameter allowing to simulate a scatterer of increasing size within mode volume without changing lateral position. This does not cause any significant damage to the microtoroid as witnessed by no observable change in the value of $\emph{Q}$ factor. As the size increases, the nanoprobe starts disturbing the already established SWMs. The evolution of SWMs and the amount of disturbance applied to $\omega^{\mp}$ and $\gamma^{\mp}$ depend on the size and location of the second probe relative to the first.
%% figure 1
%% figure 2
\begin{figure}[htbp]
\centering\includegraphics[width=8.5cm]{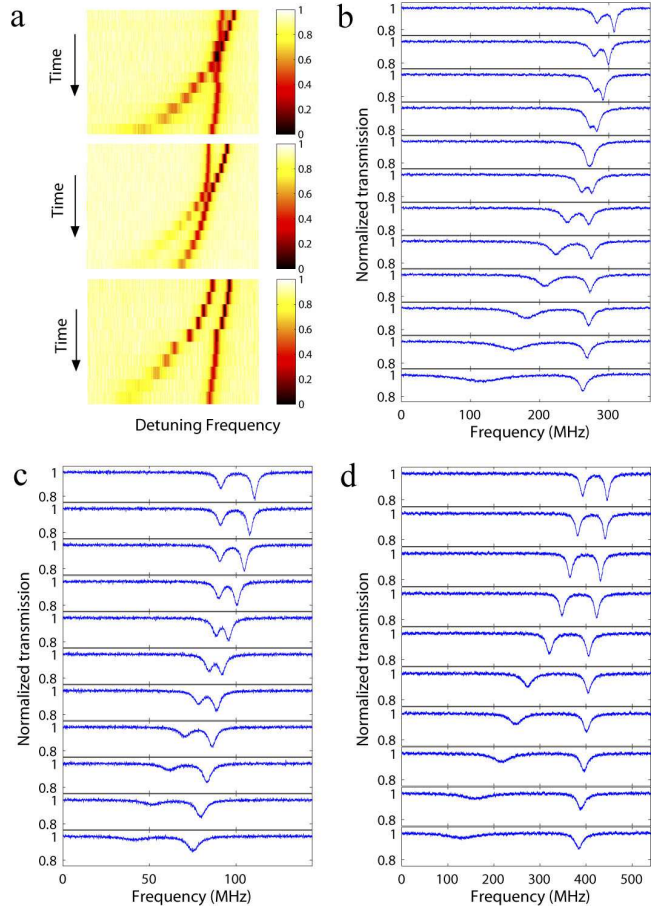}
\caption{(a) Intensity graphs of mode- crossing (top), anti-crossing (middle) and shift (bottom). (b)-(d) Transmission spectra corresponding to the intensity graphs from top to bottom in (a). Increasing time corresponds to increasing size of the second nanotip.}\label{fig2}
\end{figure}

Figure \ref{fig2} shows intensity graphs and transmission spectra recorded while the size of the second probe is increased. The second probe  is used to tune the splitting between the modes and their linewidths. Crossing (Fig. \ref{fig2}(b)), anti-crossing (Fig. \ref{fig2}(c)) and shifting (Fig. \ref{fig2}(d)) of linewidths and resonance frequencies at different lateral positions of the second probe are observed.

In Fig. \ref{fig2}(b), the initial $\omega^{+}$ mode experiences red shift and linewidth broadening with increasing size of the second probe, while the $\omega^{-}$ mode is not perturbed much. At a specific size both modes coincide, i.e., $\omega^{-}=\omega^{+}$ and  $\gamma^{-}=\gamma^{+}$. Thus, a single resonance is seen in transmission spectrum. With further increase of the second scatterer's size, the modes become separated with $\omega^{-}$ mode now having a larger linewidth than  $\omega^{+}$ whose linewidth equals to the initial $\gamma^{-}$. This suggests that both frequency and linewidth crossings have occurred. At the crossing point, back-scattering into the resonator vanishes as the back-scattered fields from the two scatterers have the same strength but $\pi$-phase shift. Thus the vanishing of backward reflection coupled to the taper suggests a transition from SWM to TWM. The conditions for these to take place will become clear in the discussion of the theoretical model below. This observation implies that SWM, which limits nonclassical features of coherent matter-cavity field interaction due to the position dependence of the coupling strength, can be eliminated using external tuning with nanoprobes. This is particularly crucial in ultra-high-$\emph{Q}$ microresonators because SWMs are usually formed due to mode splitting caused by structural defects and material inhomogenety.

In Fig. \ref{fig2}(c), the second probe first disturbs $\omega^{+}$ mode with no significant disturbance to $\omega^{-}$. Thus $\omega^{+}$ experiences linewidth broadening and red shift gradually approaching to $\omega^{-}$. At a specific scatterer size, the frequency difference between the modes reduces from its initial value of 19.6 MHz to 7 MHz, and the linewidths become very close to each other. At this point, modes are strongly coupled to the scatterer and to each other. With further increase in size, $\omega^{-}$ strongly couples to the scatterer, and the modes start to repel each other leading to increased splitting. This suggests avoided-crossing of frequency and linewidth.

In Fig. \ref{fig2}(d), the second probe affects both $\omega^{-}$ and $\omega^{+}$ and induces frequency shift and linewidth broadening. The rate of change in $\omega^{-}$ is higher than that in $\omega^{+}$ suggesting that the scatterer has a greater overlap with $\omega^{-}$.

\begin{figure}[htbp]
\centering\includegraphics[width=8cm]{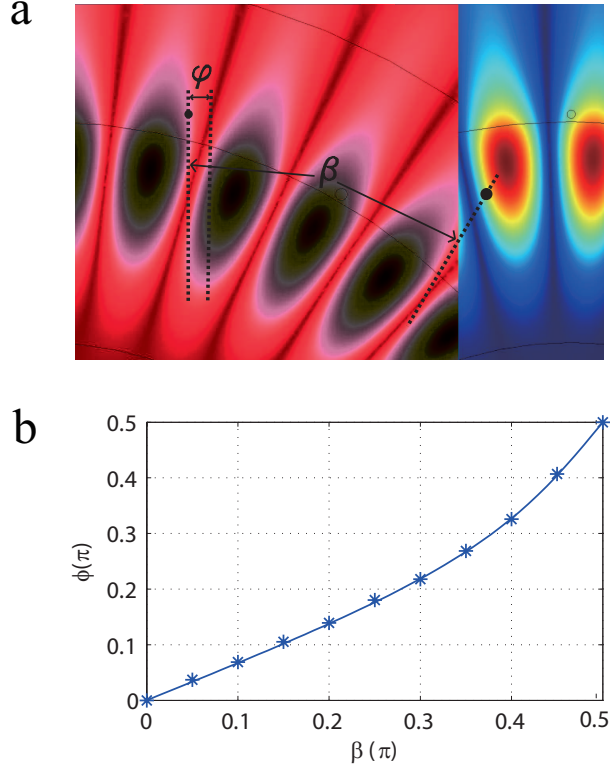}
\caption{(a) Field distribution of a standing wave mode (SWM) obtained from finite-element simulation and the definitions of $\phi$ and $\beta$. Black circles represent the position of two scatterers in the mode. (b) The relation between $\phi$ and $\beta$. Solid curve is calculated from Eq. (\ref{N02}) and $\ast$ represent the values calculated using finite-element simulations for $\chi=0.5$ and $\xi=1$.}\label{fig3}
\end{figure}

The lateral position of the second scatterer determines whether the modes will undergo a frequency crossing, anti-crossing or shifting; however, the size of the second scatterer determines the dynamics of the process. This is better understood by the following model: A single Rayleigh scatterer \cite{Mazzei} with polarizability $\alpha_1$ located at $\textbf{r}_1$ in the resonator mode volume $V$ leads to a mode splitting quantified with the coupling strength $2g_1 = -\alpha_1 f^2(\textbf{r}_1)\omega_{0}/V$ and the additional linewidth broadening $2\Gamma = \alpha_1^2f^2(\textbf{r}_1)\omega_{0}^4/3\pi c^3V$ where $c$ is the speed of light, $\omega_{0}$ is the resonance frequency before splitting, and $f^2(\textbf{r}_1)$ is the spatial variation of the intensity of the initial WGM. The resulting two SWMs have periodic spatial distributions. A single scatterer locates itself at the anti-node (node) of $\omega^{-}$ ($\omega^{+}$) with $\phi=0$ ($\pi/2$) where $\phi$ denotes the spatial phase difference between the first scatterer and the anti-node of an SWM \cite{Zhu,Mazzei}. If a second Rayleigh scatterer with polarizability $\alpha_2$ is introduced at location $\textbf{r}_2$ with a spatial phase difference of $\beta$ from the first scatterer, the already established SWMs redistribute themselves, and the amount of disturbance experienced by split modes depends on their overlap with the two scatterers (Fig. \ref{fig3}a). Subsequently, the frequency shift ($\Delta\omega^{-}=\omega^{-}-\omega_{0}$) and the linewidth broadening ($\Delta\gamma^{-}=\gamma^{-}-\gamma_{0}$) of $\omega^{-}$ with respect to the pre-scatterer resonance frequency $\omega_0$ and linewidth $\gamma_0$ become
%% equation 1,2
\begin{eqnarray}
\Delta\omega^{-}&=&2g_{1}\cos^2(\phi)+2g_{2}\cos^2(\phi-\beta)\label{system eqn1} \\
\Delta\gamma^{-}&=&2\Gamma_{1}\cos^2(\phi)+2\Gamma_{2}\cos^2(\phi-\beta)\label{system eqn2}
\end{eqnarray}
%% figure 3
where subscripts (1,2) represents the first and second scatterers. The $\cos^2(\cdot)$ terms scale the interaction strength depending on the position of the scatterer on SWMs \cite{Chantada}. Similar expressions for $\Delta\omega^{+}$ and $\Delta\gamma^{+}$ are obtained by replacing $\cos(\cdot)$ with $\sin(\cdot)$. Although we focus on two-scatterer case in this letter, this model can be extended to arbitrary number $N$ of scatterers by adding in Eqs. (\ref{system eqn1}) and (\ref{system eqn2}) the terms $2g_{i}\cos^2(\phi-\beta_i)$ and $2\Gamma_{i}\cos^2(\phi-\beta_i)$, respectively, for each of the $3\leq i\leq N$ scatterer. With the positions of the two scatterers fixed, the established orthonormal SWMs are distributed in such a way that the coupling rate between the two counter-propagating modes is maximized, in other words, the frequency splitting is maximized. It leads to:
%% Equation (3)
\begin{eqnarray}\label{N02}
\tan(2\phi)=\frac{g_{2}\sin(2\beta)}{g_{1}+g_{2}\cos(2\beta)}=\frac{\sin(2\beta)}{\chi\xi^2+\cos(2\beta)}
\end{eqnarray}
where $\xi=f(\textbf{r}_{1})/f(\textbf{r}_{2})$ and $\chi=\alpha_1/\alpha_2$ are positive real numbers. The two solutions of $\phi$ have $\pi/2$ phase difference and correspond to the two orthogonal SWMs. Verification of Eq. (\ref{N02}) is done by extensive finite-element simulations, and one example is presented in Fig. \ref{fig3}b, where finite-element simulation results match very well with the calculated values from Eq. (\ref{N02}).

Defining $\delta=\Delta\omega^{+}-\Delta\omega^{-}=\omega^{+}-\omega^{-}$ and $\varrho=\Delta\gamma^{-}-\Delta\gamma^{+}$ as the frequency and linewidth differences of the resonance modes $\omega^{-}$ and $\omega^{+}$, we find
%% Equation (4)(5)
\begin{eqnarray}
\delta=\frac{2|g_1|}{\chi\xi^2}[\chi^2\xi^4+2\chi\xi^2\cos(2\beta)+1]^{\frac{1}{2}}\label{N03a}
\end{eqnarray}
and
\begin{eqnarray}
\varrho=\frac{4\Gamma_1|g_1|}{\delta\chi^3\xi^4}[\chi^3\xi^4+\chi(1+\chi)\xi^2\cos(2\beta)+1]\label{N04}
\end{eqnarray}
from which frequency and linewidth crossings of the resonance modes can be calculated by setting $\delta=0$ and $\varrho=0$, respectively.

%% figure 4
\begin{figure}[htbp]
\centering\includegraphics[width=10cm]{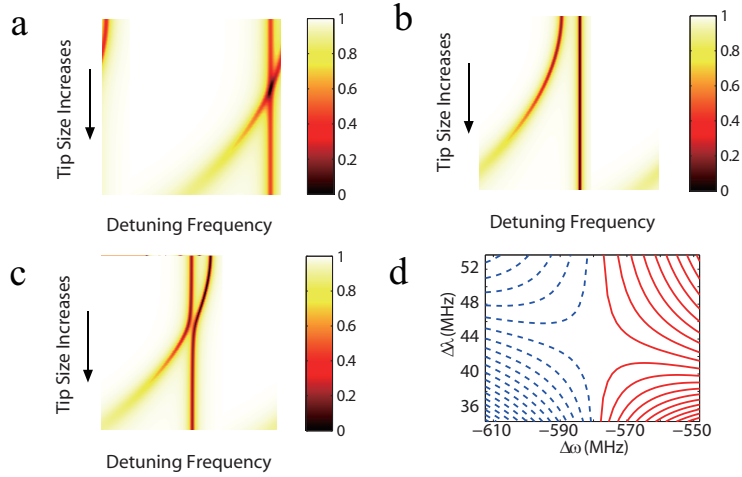}
\caption{Numerical simulations showing three unique patterns of doublet evolution for increasing size of the second nanotip at (a) $\beta=\pi/2$, (b)$\beta=0$, (c)$\beta=0.44\pi$ for $\xi=1$, (d) Resonance frequency and linewidth trajectories of the doublets with the exceptional point when $\xi$, $\chi$ and $\beta$ are varied. Dashed and solid lines correspond to the two SWMs}\label{fig4}
\end{figure}

{\it A. Behavior of the frequencies of the resonance modes:} Conditions to observe frequency crossing is found by setting $\delta=0$ which implies $\chi^2\xi^4+2\chi\xi^2\cos(2\beta)+1=0$. It is satisfied only when $\cos(2\beta)=-1$ or $\beta=\pi/2$. In the following discussions we only consider $0\leq\beta\leq\pi/2$, as $\cos(2\beta)$ is an even function and has period of $\pi$. Results of numerical simulations calculated from the model are shown in Fig.\ref{fig4}, which coincide very well with experimental observations in Fig.\ref{fig2}. Following are detailed discussions:

(i) $\beta=0$. We find $tan(2\phi)=0$, i.e., $\phi=0$, implying that both particles locate at the anti-node of $\omega^{-}$, and $\Delta\gamma^{-}$ is maximized. This leads to $\delta=2|g_1|(1+\chi^{-1}\xi^{-2})$. Thus decreasing $\chi$ increases $\delta$ by pushing $\omega^{-}$ further away from $\omega^{+}$ (Fig. \ref{fig4}(b)). Note that if the size of the second scatterer reaches above Rayleigh limit \cite{Knoll}, it may start disturbing $\omega^{+}$, too.

(ii) $\beta=\pi/2$. The second particle stays at the anti-node of $\omega^{+}$, thus increasing its size significantly affects the frequency and the linewidth of $\omega^{+}$ while its effect on $\omega^{-}$ is minimal. Then we find $\delta=2|g_1|(1-\chi^{-1}\xi^{-2})$. This implies only one frequency crossing which occurs at $\chi=\xi^{-2}$, i.e., $g_1=g_2$. For $\xi=1$, frequency crossing occurs at $\chi=1$ (Figs. \ref{fig4}(a)).

(iii) $0<\beta\leq\pi/4$. We have $0\leq\cos 2\beta<1$ (i.e., cosine is positive in the first quadrant of the unit circle) which implies that for a fixed $\beta$ in this interval, $\delta$ is always greater than zero ($\delta>0$) and it increases with decreasing $\chi$, that is with increasing $\alpha_2$. Physically, this is understood as follows. The second scatterer affects both SWMs with strengths depending on its overlap with each mode. In this interval of $\beta$, the overlap of the second scatterer with $\omega^{-}$ mode is always larger than that with $\omega^{+}$. Consequently, as $\alpha_2$ increases, $\omega^{-}$ mode is further red-shifted increasing $\delta$.

(iv) $\pi/4<\beta<\pi/2$. We have $-1\leq\cos 2\beta<0$ (i.e., cosine is negative in the second quadrant of the unit circle) implying that for a fixed $\beta$ in this interval, $\delta$ is always greater than zero ($\delta>0$); however, contrary to the case (iii) $\delta$ has a minimum at $\chi\xi^2=1/|\cos(2\beta)|>1$ with $\delta_{\rm min}=2|g_1|(1-\chi^{-2}\xi^{-4})^{1/2}>0$. The physical process is explained as follows. In this case, too, the second scatterer affects both SWMs with strengths depending on its overlap with each mode. When the size of the second scatterer is small, $\omega^{+}$ feels it strongly and undergoes frequency shift coming closer to $\omega^{-}$ as $\chi$ decreases. This changes $\phi$ and increases the overlap of the second scatterer with $\omega^{-}$ leading to their stronger interaction which consequently, red-shifts $\omega^{-}$ and helps avoid crossing $\omega^{+}$ (Fig. \ref{fig3}(c)). With a sufficiently large $\alpha_2$, $\phi>\pi/4$ will be achieved which means $\omega^{-}$ will have larger overlap with the second scatterer than the first one.

%% figure 5
\begin{figure}[htbp]
\centering\includegraphics[width=10cm]{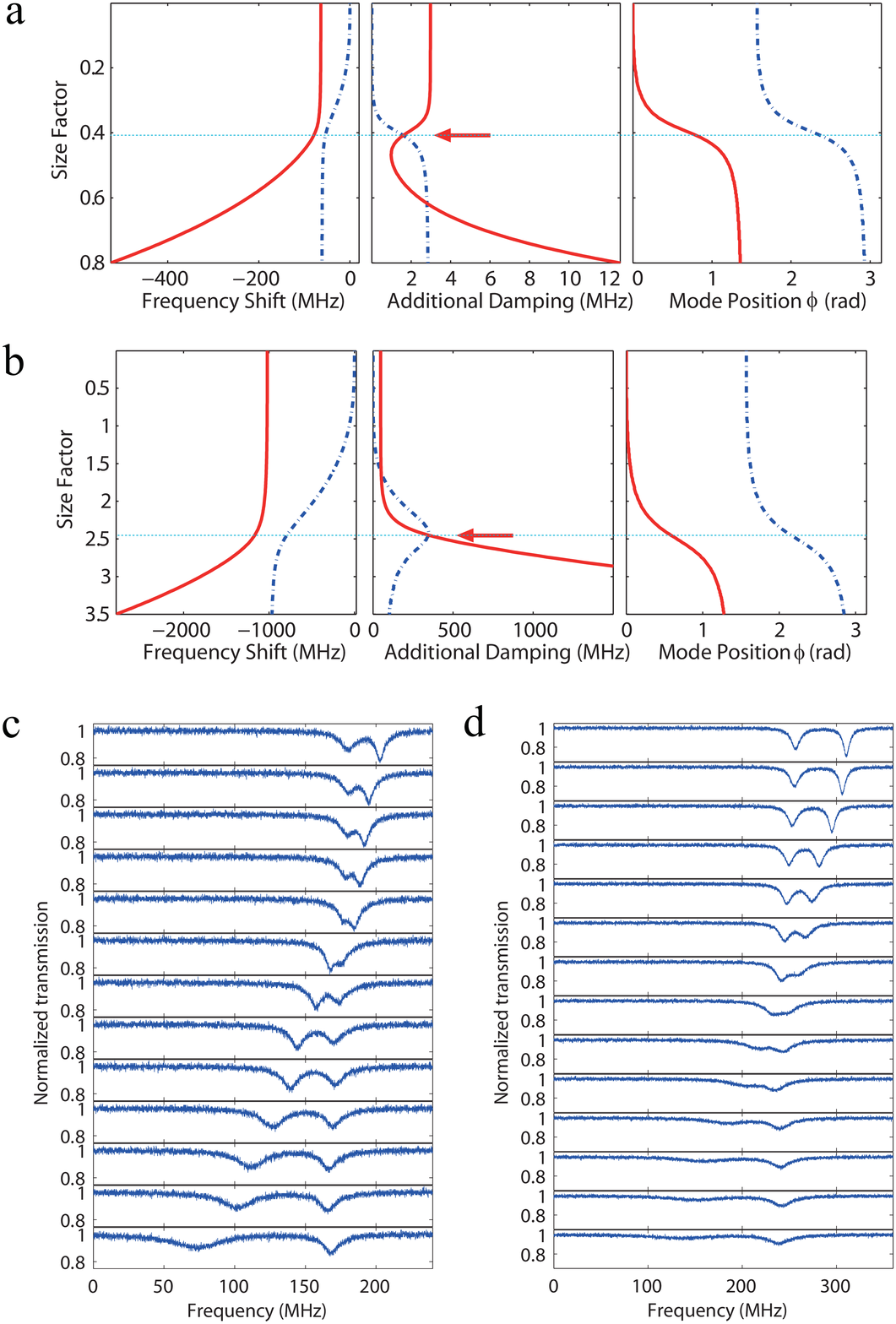}
\caption{Calculated frequency shift, linewidth broadening of the doublet and SWM position $\phi$ as a function of the ratio of second scatterer size over the first one, for (a) $\beta=0.44\pi$, $\xi=1/4$, and (b) $\beta=0.44\pi$, $\xi=4$. Dotted and solid lines correspond to the two SWMs. (c),(d) Experimental observations corresponding to (a) and (b), respectively.}\label{fig5}
\end{figure}

{\it B. Behavior of the linewidths of the resonance modes:} Setting $\varrho=0$ in Eq. (\ref{N04}), we find the condition for linewidth crossing as $1+\xi^4\chi^3+\chi(1+\chi)\xi^2\cos(2\beta)=0$ which can be satisfied only when $\cos(2\beta)<0$ or $\pi/4<\beta\leq\pi/2$, because both $\xi$ and $\chi$ are positive real numbers.

(i) $\beta=\pi/2$. We have $\phi=0$ and the two scatterers locate themselves at the anti-nodes of the each SWMs, i.e., 1st scatterer at $\omega^{-}$ and 2nd scatterer at $\omega^{+}$. Thus scatterers independently affect the two SWMs. We find that a linewidth crossing takes place at $\chi^2\xi^2=1$, which means $\Gamma_1=\Gamma_2$. In this case if we also have $\chi=\xi=1$, which gives $g_1=g_2$ and $\Gamma_1=\Gamma_2$ implying the two SWMs have identical frequency and linewidth, but have orthogonal spatial distributions. The two SWMs merge to a TWM in the direction of the initial WGM. The other directional TWM vanishes as witnessed in experiments by vanishing of backward reflection in the fiber.

(ii) $\pi/4<\beta<\pi/2$. The roots of $\varrho=0$ can be found by setting $\chi^3\xi^4+\chi(1+\chi)\xi^2\cos(2\beta)+1=0$. This is a transcendental third-order polynominal equation whose roots are too lengthy to give here. Given $\pi/4<\beta<\pi/2$, either none or two positive real roots can be found for $\chi$ at specific values of $\beta$ and $\chi$. This suggests that two linewidth crossing points may be observed. Indeed the double crossing patterns are seen in the calculated patterns shown in Fig. \ref{fig4}(a,b). In both cases, one "symmetric" linewidth crossing (indicated by arrows in Fig. \ref{fig4}(a,b)) coincides with a frequency anti-crossing. From the plot of mode position $\phi$, we see that the two SWMs "switch" distributions ($\phi$ shifts by $\pi/2$) around this point. This switching takes place around $\chi\xi^2=1$ and is the source of the symmetry of linewidth crossing. The other linewidth crossing takes place around $\chi^2\xi^2=1$, where the linewidth of one SWM changes significantly faster than that of the other one. This indicates that mode (red line in Fig. \ref{fig4}(a) and blue line in Fig. \ref{fig4}(b)) has much larger overlap with the second scatterer. Depending on whether $\xi>1$ or $\xi<1$, the "symmetric" linewidth crossing is observed before or after the other one. In experiments both scenarios were observed (Fig. \ref{fig4}(c,d)).

Moreover, in the case that no positive real roots are found for $\varrho=0$, there is no linewidth crossing although one can always find $\chi$ for specific $\xi$ and $\beta$ which minimize $\varrho$. On either side of this minimum, $\varrho$ increases implying linewidth anti-crossing. This can be explained in a similar way as the frequency anti-crossing when $\pi/4<\beta<\pi/2$.

(iii) $0\leq\beta\leq\pi/4$. In this case $\varrho\neq0$ and similar to $\delta$ in this regime, $\varrho$ increases as $\alpha_{2}$ increases. Neither crossing nor anti-crossing can be observed.

Note that the above discussed behaviors are associated with exceptional points (Fig. \ref{fig4}(d)) \cite{ExceptionalPoint} and suggest that exceptional points can be observed in a single optical resonator by precise control of external nanoprobes.

\section{Conclusion}
In conclusion, we have discussed the controlled manipulation of WGMs in a microcavity using two external nanoprobes. The theory and method described here provide guidelines for manipulating the coupling of two SWMs and probing the mode splitting related phenomena. This work suggests that many application (e.g., filters, gyroscopes, delays) which are thought to require multi-cavity configurations could be realized within a single microcavity. It also paves the ground for the discussions on tunable dual wavelength lasing with active material doped microresonators, detection of consecutively introduced nanoparticles using microresonators and studying level dynamics of interacting modes. Continuous tunability of the split WGM modes by external probes will enable to selectively amplify mechanical modes as well as help efficient cooling of mechanical modes.

\section{Acknowledgments}
The authors are grateful to MAGEEP (McDonnell Academy Global Energy and Environment Partnership) and CMI (Center for Materials Innovation) at Washington University in St. Louis  and NSF (Grant No. 0954941) for financial support. This work was performed in part at the NRF/NNIN (NSF award No. ECS-0335765) of Washington University in St. Louis.

\end{document}